\documentclass[aps,prl,twocolumn,superscriptaddress]{revtex4}

\usepackage{amsmath,amssymb,graphicx,xcolor,bm,soul}

\usepackage{url}
\usepackage[normalem]{ulem}

\usepackage{comment}
\usepackage{algorithm}

\newcommand \com[1] {\ensuremath{\mathtt{#1}}}
\usepackage{algcompatible}

\usepackage{upgreek}
\usepackage{xfrac}
\usepackage{soul}

\usepackage[bookmarks=true,
            bookmarksnumbered=false,
            bookmarksopen=true,
            breaklinks=true,
            pdfborder={0 0 0},
            final=true,
            backref=none,
            colorlinks=true,
            linkcolor=blue,
            menucolor=blue,
            citecolor=blue,
            urlcolor=blue]{hyperref}

\begin{document}

\title{MoG-VQE: Multiobjective genetic variational quantum eigensolver}

\author{D. Chivilikhin}
\affiliation{Computer Technologies Laboratory, ITMO University, St. Petersburg, 197101, Russia}

\author{A. Samarin}
\affiliation{Computer Technologies Laboratory, ITMO University, St. Petersburg, 197101, Russia}

\author{V. Ulyantsev}
\affiliation{Computer Technologies Laboratory, ITMO University, St. Petersburg, 197101, Russia}

\author{I. Iorsh}
\affiliation{Department of Nanophotonics and Metamaterials, ITMO University, St. Petersburg, 197101, Russia}

\author{A. R. Oganov}
\affiliation{Skolkovo Institute of Science and Technology, Skolkovo Innovation Center, 3 Nobel Street, Moscow 143026, Russia}

\author{O. Kyriienko}
\affiliation{Department of Physics and Astronomy, University of Exeter, Stocker Road, Exeter EX4 4QL, UK}
\affiliation{Department of Nanophotonics and Metamaterials, ITMO University, St. Petersburg, 197101, Russia}

\date{\today}

\begin{abstract}
Variational quantum eigensolver (VQE) emerged as a first practical algorithm for near-term quantum computers. Its success largely relies on the chosen variational ansatz, corresponding to a quantum circuit that prepares an approximate ground state of a Hamiltonian. Typically, it either aims to achieve high representation accuracy (at the expense of circuit depth), or uses a shallow circuit sacrificing the convergence to the exact ground state energy. Here, we propose the approach which can combine both low depth and improved precision, capitalizing on a genetically-improved ansatz for hardware-efficient VQE. Our solution, the multiobjective genetic variational quantum eigensolver (MoG-VQE), relies on multiobjective Pareto optimization, where topology of the variational ansatz is optimized using the non-dominated sorting genetic algorithm (NSGA-II). For each circuit topology, we optimize angles of single-qubit rotations using covariance matrix adaptation evolution strategy (CMA-ES) --- a derivative-free approach known to perform well for noisy black-box optimization. Our protocol allows preparing circuits that simultaneously offer high performance in terms of obtained energy precision and the number of two-qubit gates, thus trying to reach Pareto-optimal solutions. Tested for various molecules (H$_2$, H$_4$, H$_6$, BeH$_2$, LiH), we observe nearly ten-fold reduction in the two-qubit gate counts as compared to the standard hardware-efficient ansatz. For 12-qubit LiH Hamiltonian this allows reaching chemical precision already at 12 CNOTs. Consequently, the algorithm shall lead to significant growth of the ground state fidelity for near-term devices.
\end{abstract}

\maketitle

\section{Introduction}

In recent years quantum computing made a giant leap from being a purely theoretical concept into forming a full-scale quantum industry. Its major applications are in the fields of quantum chemistry and materials science~\cite{OxfordRev,ZapataRev}. A particularly promising task for quantum computing is studying low energy states of strongly-correlated molecular complexes and materials~\cite{Reiher2016,Babbush2019}. By using quantum phase estimation and Hamiltonian simulation techniques~\cite{Lloyd1996,Aspuru-Guzik2005}, one can perform calculations with exponentially improved scaling as compared to classical computational methods. However, this typically requires large auxiliary qubit registers and controlled multi-qubit operations. As modern quantum hardware is prone to noise, the implementation of deep quantum circuits becomes impractical unless quantum-error correction is used. The viable alternative for noisy intermediate scale quantum (NISQ) devices~\cite{Preskill2018} corresponds to hybrid quantum-classical approaches that rely on the variational ground state search.
\begin{figure*}[t!]
\includegraphics[width=0.75\textwidth]{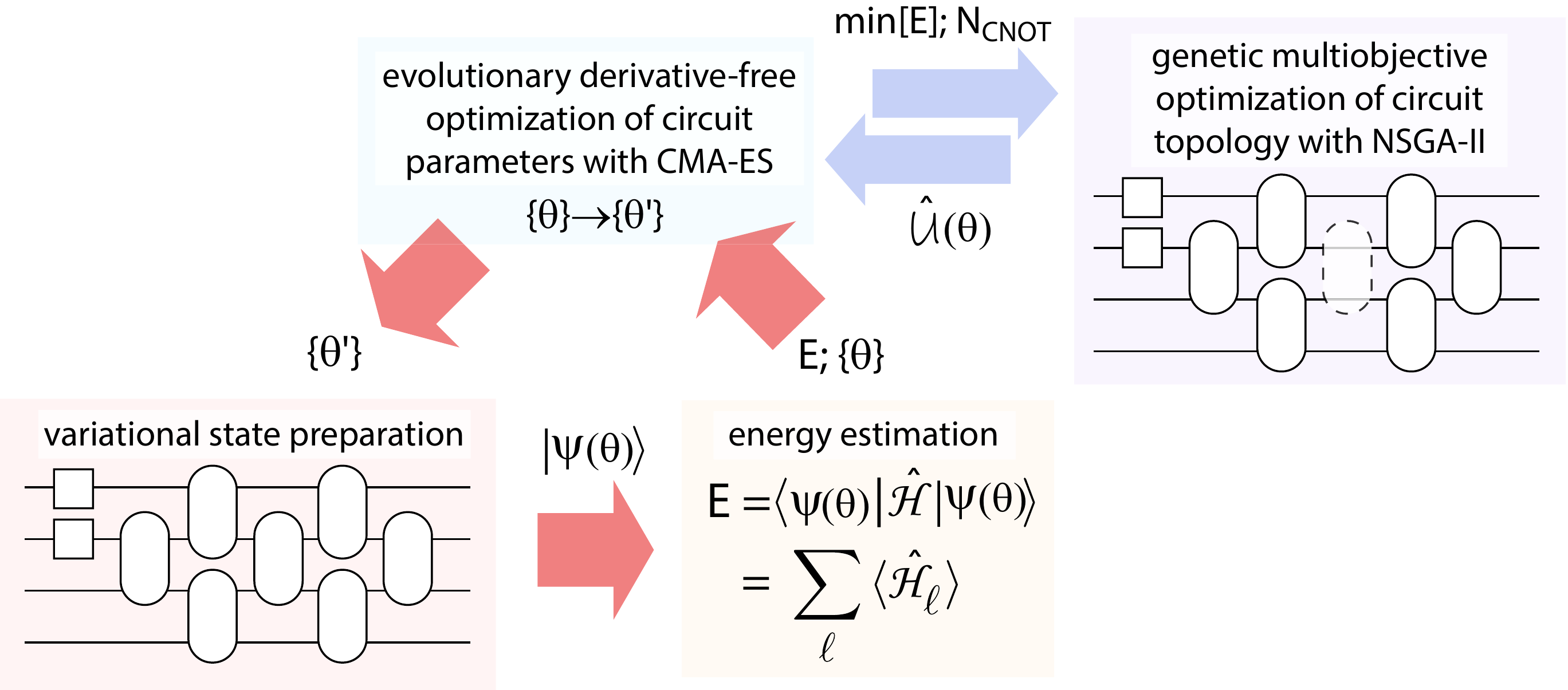}
\caption{\textbf{Multiobjective genetic VQE workflow}. The goal of the algorithm is to optimize a circuit ansatz $\hat{\mathcal{U}}(\bm{\theta})$ for preparing the ground state of the Hamiltonian $\hat{\mathcal{H}}$. First, a quantum circuit layout is chosen using a multiobjective genetic algorithm (left block, NSGA-II), where the number of CNOT operators is minimized alongside the energy $\langle \psi(\bm{\theta})| \hat{\mathcal{H}} |\psi(\bm{\theta})\rangle$. Next, for the chosen scheme variational angles $\{\bm{\theta}\}$ are defined using an evolutionary optimizer (middle block, CMA-ES). Function evaluations are performed by the quantum hardware, which in our case is simulated classically.}
\label{fig:algorithm}
\end{figure*}

The first example of a hybrid quantum-classical algorithm is the variational quantum eigensolver~\cite{Peruzzo2014,McClean2016,Romero2019}, which combines the quantum state preparation with a classical optimization procedure. The standard workflow can be sketched as: 1) start from the initial state $|\psi_0\rangle$ as a simple product state, being a closest match to the ground state of the Hamiltonian $\hat{\mathcal{H}}$; 2) construct a unitary operator $\hat{\mathcal{U}}(\bm{\theta})$ as a variational ansatz parametrized by angles $\{ \bm{\theta} \}$; 3) prepare an approximation to the ground state as $|\psi(\bm{\theta})\rangle = \hat{\mathcal{U}}(\bm{\theta}) |\psi_0\rangle$; 4) measure the expected energy as $E = \langle \psi(\bm{\theta}) | \hat{\mathcal{H}} | \psi(\bm{\theta}) \rangle$; 5) update parameters using a classical optimization procedure, and repeat steps 3, 4, and 5 until energy minimum is reached.

By now variational quantum algorithms are represented by the family of protocols, which can be tentatively divided into chemically-inspired and hardware-efficient approaches (see recent reviews~\cite{OxfordRev} and~\cite{ZapataRev} for the detailed discussion). For the chemically-inspired methods a unitary version of the coupled cluster ansatz is considered~\cite{Romero2019}, which is routinely used in quantum chemistry, and allows converging close to the system ground state. At the same time, the chemically-motivated ansatz typically relies on implementing a large number of Pauli operator strings (originating from fermions-to-qubits mapping). This makes it an expensive ansatz for near-term devices. The approach was used for experimental demonstrations of the ground state energy estimation ranging from the simple molecular hydrogen molecule~\cite{OMalley2016} to recent state-of-the-art VQE for H$_2$O molecule with an ion-trap processor~\cite{IonQ2019}. The hardware-efficient approach represents a chemistry-agnostic ansatz that is formed by layers of entangling operations (usually taken as CNOT gates) and rotations with angles~$\{\bm{\theta}\}$, such that at increased number of layers a ground state can be prepared. First demonstrated experimentally for two, four, and six-qubit systems~\cite{Kandala2017,Ganzhorn2018,Hempel2018}, the approach offers resource-efficient simulation at the expense of growth of variational parameters number. At the same time, the convergence with the number of layers is unknown, and strongly-correlated molecules generally require a large number of layers and two-qubit operations. Finally, a different approach based on Givens rotations was used to prepare variationally Hartree-Fock states of hydrogen chains and study isomerization of diazene~\cite{Arute2020}.  

Various improvements upon the basic VQE workflow were proposed. For instance, the state-of-the-art optimization strategies based on imaginary-time evolution and natural gradient techniques were developed~\cite{McArdle2018,Endo2018,Stokes2020,Yamamoto2019,Balint2019}. Symmetry constraints can be introduced into the ansatz~\cite{Ryabinkin2019,Gard2019}, thus simplifying the circuit and lowering the gate count. A mixed variational approach with elements of quantum phase estimation allowed accelerating  VQE into the genuine ground state direction~\cite{Riverlane2019}. Derivative-based methods were considered, where the use of analytic derivatives for unitary circuits was shown to be highly beneficial~\cite{Mitarai2019,OBrien2019}. Huge importance of the form of the variational cost-function was noted~\cite{Cerezo2020}. Time-grid methods based on state overlap measurement recently gained attention~\cite{Kyriienko2019,Stair2020,Parrish2019,Wei2020}, also using a variational ansatz as a sum of unitaries~\cite{Huggins2019}. State-of-the-art strategies with compressed pools of variational operators were successfully applied to improve variation~\cite{Grimsley2019,Tang2019,Ryabinkin2019b,Herasymenko2019}. Finally, reinforcement learning was employed to enhance quantum eigensolver workflow in a semi-automated fashion \cite{Lamata2020}.

In this paper we tackle the problem of finding an optimal variational ansatz for chemistry problems using a system-agnostic and \emph{hardware-efficient} ansatz, pushing it to the ultimate limit. The proposed Multiobjective Genetic Variational Quantum Eigensolver~(MoG-VQE) performs multiobjective optimization of the ansatz circuit, where an original block-based ansatz circuit is genetically modified to yield quasi-optimal solutions in terms of energy precision and number of entangling two-qubit operations. Considering particular examples of 8- and 12-qubit Hamiltonians corresponding to BeH$_2$, H$_4$, and LiH molecules, we show that the number of CNOT operations can be reduced by an order of magnitude, being as low as 12 CNOTs for chemically precise preparation of 12-qubit LiH ground state. 
\begin{figure*}
\includegraphics[width=1.\linewidth]{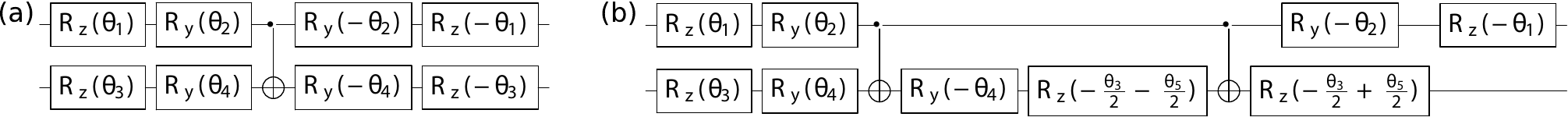}
\caption{\textbf{Circuit blocks.} (a) Generalized CNOT block, where a CNOT gate is sandwiched between single-qubit rotations to choose an optimal basis. The middle gate represents a standard CNOT controlled by the top qubit, and single-qubit rotation gates are defined as $R_z(\theta) = \exp\left(-i\frac{\theta}{2} \hat{Z} \right)$, $R_y(\theta) = \exp\left(-i\frac{\theta}{2} \hat{Y} \right)$, $R_x(\theta) = \exp\left(-i\frac{\theta}{2} \hat{X} \right)$. (b) Generalized two-qubit gate block, which uses two CNOTs and rotations defined by five angles. }
\label{fig:blocks}
\end{figure*}

\section{The MoG-VQE algorithm}

First, we recall the standard hardware-efficient VQE approach. It was proposed and used by the IBM Quantum group to find experimentally the ground state energy of small molecules with a superconducting quantum processor~\cite{Kandala2017}. The hardware-efficient VQE uses a layered structure of a variational ansatz. The zeroth layer corresponds to choosing the basis by variation of optimal angles (single qubit rotations) on top of a trivial binary initial state. Next, the layer of CNOTs connecting all qubits in pairwise fashion is applied, followed by the set of single-qubit rotations (generated by $\hat{X}$ and $\hat{Z}$ Pauli operators) with variable angles. The procedure continues by growing the number of layers (depth $p$), and typically a large value of $p$ (of the order or larger than number of qubits) is required to reach chemical precision of complex molecules.

In this work we adopt a different approach which assumes that not all qubits shall be equally correlated, and excessive entangling operations may make optimization inefficient. This is also in line with recent results showing the connection between entanglement production and efficiency of convergence~\cite{Woitzik2020}. We propose to prepare the unitary $\hat{\mathcal{U}} (\bm{\theta})$ as a variable scheme composed of blocks, with topology being decided in the process of multiobjective genetic search. From one side, the block structure resembles the previously described hardware-efficient ansatz (HEA), where only parts from each layer are utilized. From the other side, the similar block structure can be shown to represent a matrix product state ansatz. The high-level description of the approach is presented in Fig.~\ref{fig:algorithm}. This corresponds to sequential optimization of both energy of the system and the variational ansatz itself.

For optimization of the VQE circuit topology in this work we propose to use a (multiobjective) genetic algorithm~(GA)~\cite{mitchell} from the family of evolutionary algorithms~-- general-purpose heuristics based on principles of biological evolution used for optimizing arbitrary functions. The optimization problem solved by GA is described by one or several optimization functions called \emph{fitness functions}. Genetic algorithms operate with a population (set) of one or several \emph{individuals}~-- representations for solutions of the optimization problem. On each iteration (also called \emph{generation}) a GA performs \emph{variation} and \emph{selection} on the current population to determine the individuals that will advance to the next \emph{generation}. During variation individuals from the current population are modified with \emph{mutation} and/or \emph{crossover} operators. A mutation operator acts on one individual and introduces a relatively small change to it.
A crossover operator typically acts on two \emph{parent} individuals and produces also two individuals, which have parts of parent individuals. The selection procedure determines individuals which form the next population.

In MoG-VQE we use GAs for two purposes: 1) optimizing the VQE circuit topology, and 2) optimizing angles for single-qubit gates for a fixed circuit topology. When optimizing the VQE scheme topology we consider two objective (fitness) functions, namely, the energy and the number of two-qubit gates (CNOTs), both to be minimized. To do it in the most efficient way we use the multiobjective genetic algorithm NSGA-II~\cite{nsga} that allows optimizing several objective functions simultaneously.
As a result, NSGA-II tries to approximate the set of Pareto-optimal solutions, i.e. such solutions that no fitness function can be improved without degrading some other fitness functions. Each circuit (individual of the GA) is described as a list of gates, each gate associated to one or two qubits depending on its type. Instead of letting the GA to compose circuits from elementary quantum gates, we make it operate at the level of gate \emph{blocks} composed of one or several CNOTs and several rotation operators (see Fig.~\ref{fig:blocks} and the discussion below).

\textit{Gate blocks design}.
To ensure the convergence to the ground state, gate blocks shall be chosen such that effectively each state in the Hilbert space can be reached. For this, various strategies can be employed. First, let us consider the CNOT gate, where each qubit can be rotated by an arbitrary angle. The corresponding operation can be found in~\cite{Iten2016} representing a generalized CNOT, shown as a circuit in Fig.~\ref{fig:blocks}(a). This circuit implements a CNOT in both directions efficiently (by effectively changing control and target qubits), and also allows conversion of a CNOT to a CZ gate. The decomposition requires 4 variational angles (initially set randomly, though may be initialized using an intelligent guess), and can simplify the variational procedure. However, it cannot be reduced to an identity matrix.

Second, we can use a more general block that includes two CNOT gates. The idea comes from the arbitrary control operation~\cite{NielsenChuang}. Additionally, we consider an arbitrary rotation for the top qubit, such that we are not restricted to a certain control state. The resulting circuit is shown in Fig. \ref{fig:blocks}(b). It has 5 variational angles and 9 single-qubit gates total. For instance, it should allow generation of a $\sqrt{\text{SWAP}}$ gate from CNOTs. Importantly, we can choose a non-trivial set of angles as $\theta_4= \pi, \theta_5 = 2\pi - \theta_3$, and assign $\theta_{1,2,3}$ randomly from the $(-\pi,\pi]$ interval. In this case, the circuit turns into an identity matrix, and thus will utilize the previously optimized state~\cite{Grant2019}. We envisage there are many other choices that can improve performance, and few ideas are mentioned in the Discussion section.

\textit{Circuit initialization}. For the zeroth population of NSGA-II we initialize each circuit in the following way. With a probability of $1/2$, a circuit is constructed as a check-board of blocks on odd/even sublattices. This aims to  correlate and entangle all qubits in the lattice. The procedure thus adds $N_{\text{init}} = N - 1$ blocks, where $N$ is the number of qubits.
Alternatively, also with probability $1/2$, we use the following circuit initialization procedure. We select the number of blocks $N_{\text{init}}$ to be added uniformly at random from the interval $[N, 4N]$, and then add $N_{\text{init}}$ blocks targeting randomly selected pairs of qubits. 

This combined strategy provides necessary diversity of the initial population of the GA, and also deals with the issue of the a priori unknown complexity of the problem. 
Indeed, different circuit sizes are needed for different Hamiltonians, and, for example, if initialized with an insufficient circuit size, MoG-VQE will require many iterations to reach the needed size.
By providing a diverse (in terms of size and structure) set of circuits in initialization we facilitate faster convergence to the ground state.

\textit{Circuit variation}. To vary the circuit topology we use a simple mutation operator. It uses two basic operations: insertion of a block at a random position in the circuit, and deletion of a block from a random position in a circuit. The mutation operator selects one of three operations to perform according to a random weighted choice (the larger the weight, the larger the probability): 1) insert a block with weight $w_{\text{insert}} = 2.0$; 2) delete a block with weight $w_{\text{delete}} = 1.0$; 3) ``large-scale'' mutation with weight $w_{\text{big}} = 0.25$, performing 10 insertions/deletions. The latter operation is selected with low probability and is intended for escaping from local minima.
As for crossover, initial experiments demonstrated that simple one-point and two-point crossover operations~\cite{mitchell} for lists of circuit blocks are inefficient, so we omit using this operator in the current version of the algorithm.
For selection we use the tournament selection operator proposed in~\cite{nsga} for multiobjective evolutionary optimization, which is based on non-dominance of solutions and crowding distance.
We note that so far no physical topology for a quantum hardware is specified, and new blocks are inserted to act on chosen qubits $i$ and $k$. Later, this can be rewritten in the optimal form targeting a specific device, or encoding actual topology already when performing NSGA-II. 
\begin{figure*}[t]
    \includegraphics[width=1.\linewidth]{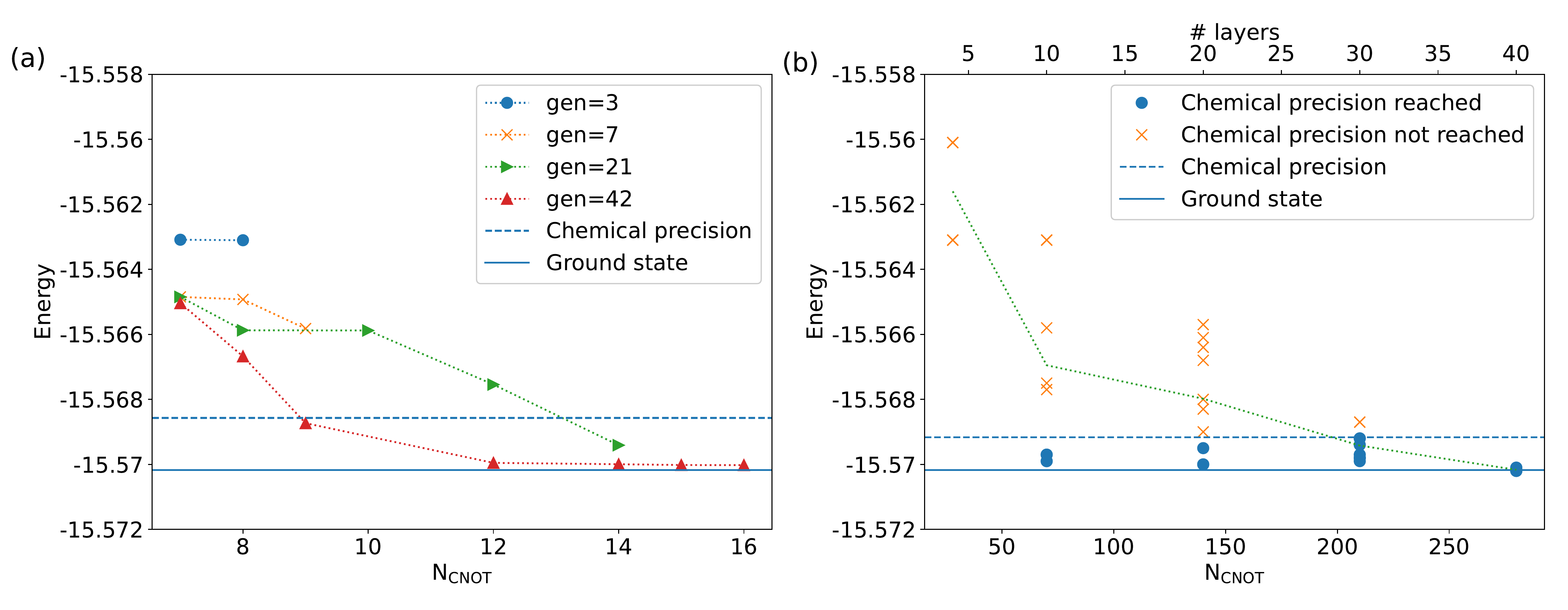}
    \caption{\textbf{MoG-VQE vs. HEA for BeH$_2$.} Results of numerical simulations are shown for beryllium hydride molecule BeH$_2$ at the equilibrium bond distance of $1.33$ {\AA}ngstrom. (a) Pareto fronts generated by the proposed MoG-VQE approach for different generations of NSGA-II optimization. Here each Pareto front corresponds to a quasi-optimal trade-off solution in coordinates of CNOT number and energy, where both objectives cannot be simultaneously improved. The blue solid line corresponds to the ground state energy from exact diagonalization of the BeH$_2$ Hamiltonian, and the blue dotted line shows a minimal chemically accurate solution. Chemical precision can be reached with a minimal number of $9$~CNOTs. (b) We compare results of MoG-VQE with genetically modified ansatz to the hardware-efficient strategy with fixed number of layers. Energies reached by HEA are shown for increasing numbers of layers (top scale), and corresponding numbers of CNOT gates are shown at the bottom. Chemical precision can be reached for the minimal number of $70$~CNOTs.}
    \label{fig:beh2plots}
\end{figure*}

\textit{Objective functions evaluation}. As mentioned above, we use two objective functions~-- the energy of the system and the number of CNOTs in the circuit. Calculation of the second objective function is trivial. However, to calculate the first objective function (expected value of energy), we need to optimize variational angles for the respective gates in the circuit. Traditionally, optimization of angles is done using classical methods based on numerical differentiation such as the BFGS algorithm~\cite{bfgs} (suffers from noise when run on physical devices) or gradient-free methods that include the Nelder-Mead simplex method~\cite{nelder-mead} and COBYLA~\cite{Romero2019}. Recent advances include use of stochastic gradient descent optimization and, in particular, Adam~\cite{adam} being widely popular in machine learning (see comparison in \cite{Kubler2020}).
In this work we propose to use an evolutionary algorithm, namely the Covariance-Matrix Adaptation Evolutionary Strategy~(CMA-ES)~\cite{cmaes}.
CMA-ES is a heuristic derivative-free method for non-linear or non-convex continuous optimization problems that has been shown to be effective for hard problem instances.
Each time we need to evaluate the energy for a particular circuit topology, we launch CMA-ES to optimize angles of single-qubit rotation gates. During a single run, CMA-ES starts from random angles for single-qubit gates and varies them to optimize the energy. For each valuation of angles the energy is calculated using a Hamiltonian averaging procedure implemented on a quantum processor (or its simulator).
The energy found by a single run of CMA-ES is taken as the optimal energy for the circuit. This is, of course, a heuristic, since due to the stochastic nature of CMA-ES, different optimization runs may result in different minimal energy values. For a trade-off between evaluation time and accuracy, we can also select the number of independent optimization runs of CMA-ES: the minimal energy found in these runs will be taken as the optimal energy for the circuit.

A pseudocode of the proposed MoG-VQE approach is shown in Algorithm~\ref{alg:alg}.
The algorithm starts with initializing the population of potential solutions (circuits) $P$ according to the procedure described above.
Then, in each generation a new population $P'$ of solutions is formed by mutating each individual and optimizing rotation angles of each circuit using CMA-ES.
Note that the innter for-loop is executed in parallel. 
The algorithm terminates if some termination criterion (e.g., solution with chemical precision is found, or a given number of generations is reached) is satisfied.

\begin{algorithm}
\begin{algorithmic} 
    \REQUIRE number of qubits $N$
    \REQUIRE population size $n$
    \REQUIRE Hamiltonian $\hat{\mathcal{H}}$
    \STATE $P \gets \com{initPopulation}(n, N)$
    \FOR{$gen \gets 0 \; \mathbf{to} \; \infty$} 
        \STATE $P' \gets \{\}$
        \FOR{$c \in P$} 
            \STATE $m \gets \com{mutate}(c)$
            \STATE $m.{\text{energy}} \gets \com{CMA\text{-}ES}(m, \hat{\mathcal{H}})$
            \STATE $m.{N_\text{CNOT}} \gets \com{countCNOTs}(m)$
            \STATE $P' \gets P' \cup \{m\}$
        \ENDFOR
        \STATE $P \gets \com{NSGAselection}(P, P')$
        \IF {\com{terminate}(P, gen)} 
            \STATE \textbf{return} $P$
        \ENDIF
    \ENDFOR
\end{algorithmic}
\caption{Pseudocode of the proposed MoG-VQE approach.}
\label{alg:alg}
\end{algorithm}

\textit{Software implementation}.
Circuit topology optimization is implemented in {\sffamily{}Python} using the {\sffamily{}DEAP} library~\cite{deap} for NSGA-II implementation. Optimization of angles for rotations is implemented in {\sffamily{}C++} language using {\sffamily{}libcmaes}~\cite{libcmaes} library for CMA-ES. To simulate quantum circuits efficiently it is crucial to choose a fast software simulator of a quantum processor. We have benchmarked several packages for this task, including the statevector simulators in {\sffamily{}QuTiP}~\cite{Qutip} and {\sffamily{}Qiskit}~\cite{Qiskit} packages for {\sffamily{}Python}, and the high-performance Quantum Exact Simulation Toolkit ({\sffamily{}QueST})~\cite{QuEST} written in {\sffamily{}C++}. We find {\sffamily{}QueST} to have better performance, in many ways due to efficient {\sffamily{}C++}-based operations and thus use it for MoG-VQE, with {\sffamily{}Python} bindings to interact with the circuit topology optimizer. Circuits are visualized and compiled in a {\sffamily{}.qasm} files using {\sffamily{}Qiskit} package from IBM Quantum Experience~\cite{Qiskit}.


\section{Results}
We apply the developed multiobjective genetic VQE to various problems in quantum chemistry. For this, we consider a set of molecular Hamiltonians as generated by {\sffamily{}Psi4} quantum chemistry package~\cite{psi4}, and mapped into qubit Hamiltonian via Jordan-Wigner transformation using {\sffamily{}OpenFermion} package~\cite{OpenFermion}. We also use the problem set from QunaSys competition~\cite{QunaSys} (generated with {\sffamily{}PySCF} package~\cite{PySCF}). In this set Hamiltonians for H$_n$ ($n=4,6$) and LiH molecules in various geometries are considered at different bond distances. In each case the STO-3G minimal basis set for spin singlets was used.

In all computational experiments with the proposed approach we used the following settings. For NSGA-II we used a population size of $64$ and mutation probability of $1.00$. The basic CMA-ES version implemented in~\cite{libcmaes} was used with parameters: tolerance~=~$10^{-5}$, $\sigma$~=~0.5. Experiments were run using 16 cores of a 64-core AMD Opteron 6380 processor @ 2.5 GHz: after each circuit variation step in NSGA-II, energy is optimized with CMA-ES in parallel for batches of 16 circuits. For all plotted results blocks of the first type [Fig.~\ref{fig:blocks}(a)] are used, while we comment on second block choice in the text. In each case the initial state is chosen as a Hartee-Fock state, that for relatively small molecules corresponds to the product state having the largest overlap with the genuine many-body Hamiltonian ground state.

We compare the developed MoG-VQE approach with HEA~\cite{Kandala2017} with different number of layers $p$. In the latter the circuit topology is fixed during VQE and represents layers of CNOTs placed on neighbouring qubits followed by rotations, repeated $p$ times. To make a valid comparison, when optimizing rotation angles for HEA we initially used the same algorithm as in our approach, namely, CMA-ES. However, since the depth of resulting circuits (and thus the number of optimization parameters) for HEA is much larger than for our approach, the original CMA-ES performs poorly. This is mostly due to the use of operations that have a quadratic time complexity dependence from the number of optimization parameters. Therefore, below we report results in which HEA parameters were optimized using a variation of CMA-ES called sep-CMA-ES~\cite{sepCMAES}, in which each iteration has linear time complexity.
\begin{figure*}
    \includegraphics[width=0.9\linewidth]{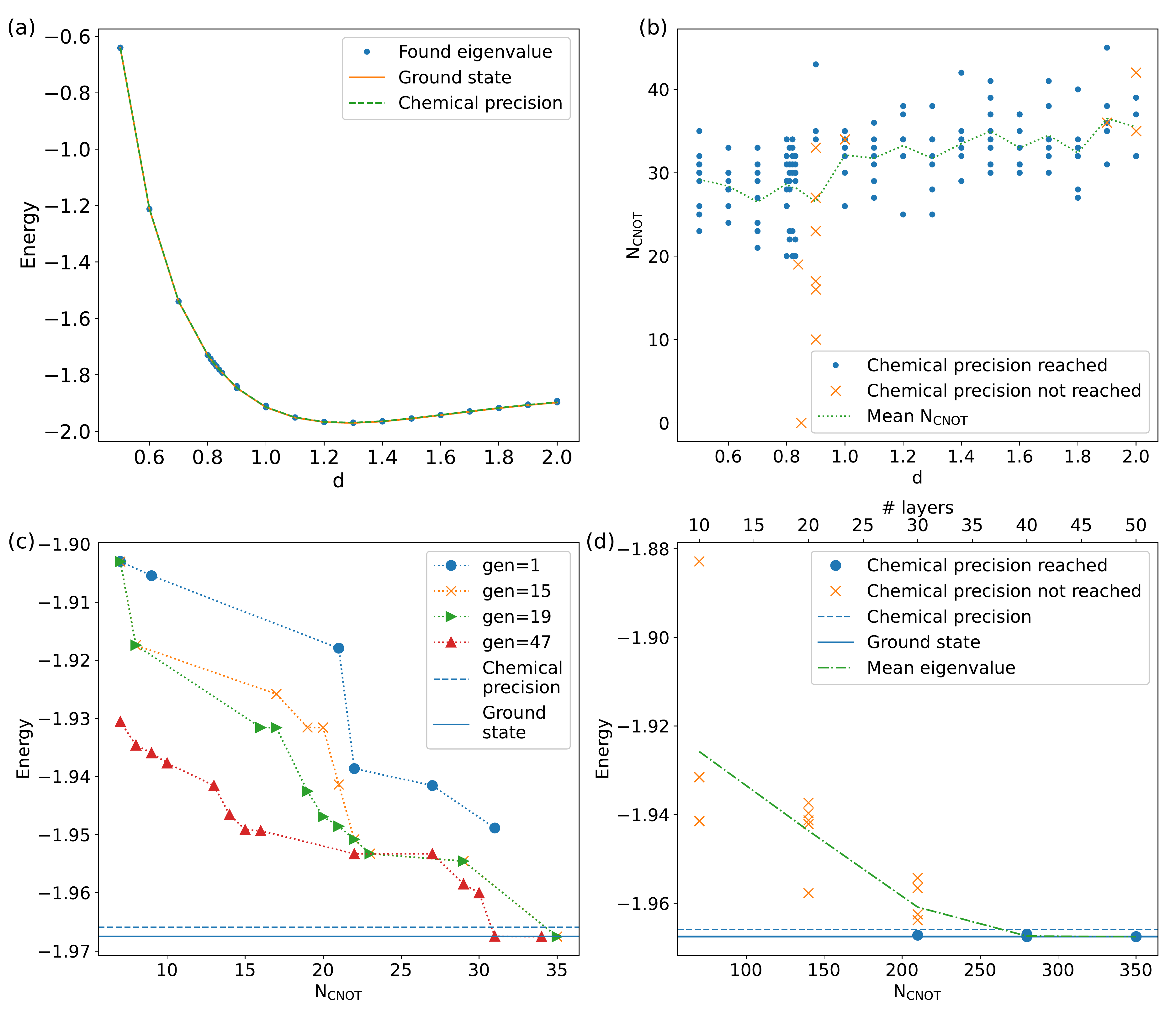}
    \caption{\textbf{MoG-VQE vs. HEA for hydrogen chains.} We consider the H$_4$ molecule in the open chain (line) geometry, where the bond distance $d$ (measured in \AA) between atoms is uniform. (a) Potential energy curve for H$_4$ molecule shown as a function of bond distance $d$. Results of MoG-VQE are depicted by blue dots, and are shown to recover genuine ground state energies obtained from exact diagonalization. (b) Two-qubit gate counts for MoG-VQE circuits needed to prepare approximate ground states at different $d$. Circuits for runs that did not provide chemical accuracy are shown by crosses. The dashed curve represents the mean CNOT number for different runs, and serves as a guide to an eye. (c) Pareto fronts plotted for energy and $N_{\mathrm{CNOT}}$ objective functions, and displayed for different generations. Here, we consider H$_4$ at bond distance of $d=1.20$ {\AA}ngstrom. (d) Results for the HEA VQE simulations of H$_4$ ($d=1.20$~\AA). Minimal energy shown as a function of numbers of layers (top), and corresponding number of CNOT operations.}
    \label{fig:h4plots}
\end{figure*}

\textit{Beryllium hydride BeH$_2$ (8 qubits).} As a first example, we consider BeH$_2$ at the bond distance of $1.33$~{\AA}ngstrom, being a stable configuration. The Hamiltonian is obtained using different active space parameters, and is encoded with $N=8$ qubits~\cite{Kyriienko2019}. We perform multiobjective VQE, and compare it to the hardware efficient ansatz proposed in~\cite{Kandala2017}. The results are shown in Fig.~\ref{fig:beh2plots}. 
\begin{figure*}
    \includegraphics[width=1.\linewidth]{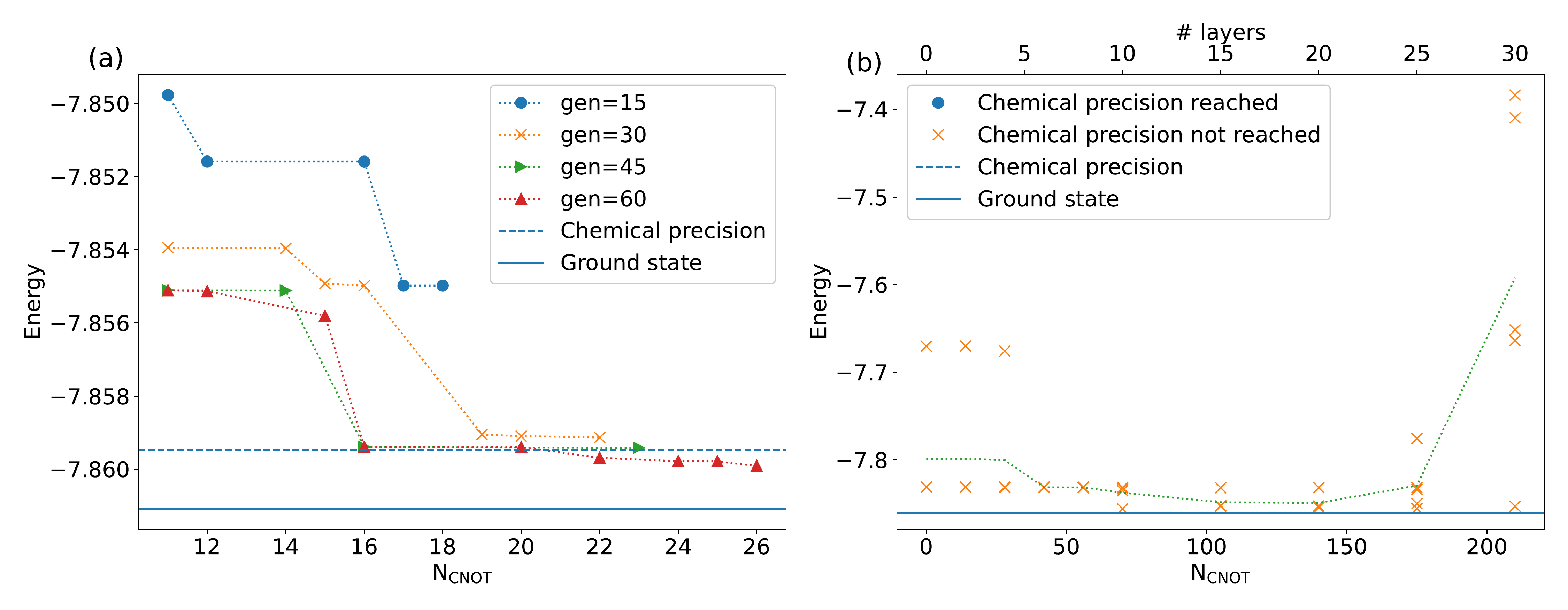}
    \caption{\textbf{MoG-VQE vs. HEA for LiH.} (a) Energy-$N_{\mathrm{CNOT}}$ Pareto fronts for multiobjective VQE for lithium hydride at bond distance $d=2.00$~\AA. After sufficient optimization (for $\mathrm{gen}>30$) shallow circuits with $N_{\mathrm{CNOT}} = 16$ are found. (b) Comparison of the proposed approach with the hardware-efficient ansatz.}
    \label{fig:LiH}
\end{figure*}

Here and below, the bottom solid line corresponds to the full configurational interaction (FCI) ground state energy from exact diagonalization, and the top dashed line depicts FCI energy within chemical accuracy, which allows for $10^{-3}$ Hartree error at room temperature. The number of CNOT operations comes directly from the number of blocks, where single-qubit operations supplement each two-qubit gate. 
In Fig.~\ref{fig:beh2plots} we plot the minimal obtained energy as a function of $N_{\mathrm{CNOT}}$ obtained by MoG-VQE, and provide the comparison to HEA.  Fig.~\ref{fig:beh2plots}(a) shows Pareto fronts of solutions that account for the trade-off between energy minimization and number of CNOTs (circuit depth). Progressing from inefficient ansatz circuits of early generations, we see that at the 21st generation the GA reaches chemical precision with 14 CNOTs. As we continue to optimize variational circuits, Pareto fronts are bettered, with both energy precision and circuit depth being improved (gen=42 curve), shifting to the left bottom corner. The circuit containing as low as 9 CNOTs can be constructed in the way that a chemically-accurate ground state of 8-qubit BeH$_2$ can be prepared. Generally we observe that forty to sixty generations are required to obtain quasi-optimal circuits that allow preparing low-energy states and are resource-efficient.

Additionally, we test the quality of obtained solutions tracking the total magnetization and parity of prepared states. As the system corresponds to interacting fermions in the absence of a magnetic field, both charge conservation and time-reversal symmetry must be satisfied. This, for instance, is captured by the ansatz choice that by construction does not change magnetization and parity for any physical state~\cite{Gard2019}. While our ansatz does not obey required symmetries explicitly, we find that magnetization and parity of variationally prepared states differ from the initial ones only in the third digit after the decimal point. This allows us to confirm that MoG-VQE indeed prepares physical ground state solutions. We also note that for some cases symmetries can be encoded in additional fitness functions, enjoying the power of multiobjective optimization.

Next, we perform calculations for the hardware-efficient ansatz described above. For HEA we repeat numerical experiments 5 times for each circuit depth. In Fig.~\ref{fig:beh2plots}(b) we show minimal HEA VQE energy as a function of the number of layers, or equivalently the number of CNOTs (each point for the same $N_{\mathrm{CNOT}}$ corresponds to a separate computational experiment). We find that the minimal number of layers for attaining chemical precision is 10 (70 CNOTs), while to reach chemical precision with larger probability 30 layers (210 CNOTs) are needed.

Studying the efficiency of different blocks we discover that the block of first type [Fig.~\ref{fig:blocks}(a)] has better performance, and corresponds to our top choice for MoG-VQE. The reason is its versatility, as several blocks put in a series can represent other circuits at the same cost as a single two-qubit operation. On the contrary, we find that VQE with blocks of the second type [Fig.~\ref{fig:blocks}(b)] has weaker performance (not shown). While being rather general, training of the circuit network with 5-parameter blocks becomes challenging while also restricted by $N_{\mathrm{CNOT}}$ set as one of the objective functions.

\textit{Hydrogen chain H$_4$ in a line geometry (8 qubits).} Next, we consider a highly correlated molecule corresponding to the hydrogen chain H$_4$ (line geometry), where equal bond distance $d$ and open boundary are taken~\cite{QunaSys}. This corresponds to $N=8$ qubit Hamiltonian in STO-3G basis. Bond distances from $0.50$~\AA~to $2.00$~\AA~are considered. Note that for bond distances of $1.3$~\AA  ~and more, the initial energy (Hartree-Fock solution) and FCI energy are largely separated, making it a difficult molecule to study. 

Results of MoG-VQE experiments for the hydrogen chain are shown in Fig.~\ref{fig:h4plots}. In panel Fig.~\ref{fig:h4plots}(a) we present a potential energy curve, showing that MoG-VQE allows accurate reproduction of the FCI ground state obtained by exact diagonalization of the molecular Hamiltonian. In Fig.~\ref{fig:h4plots}(b) we present two-qubit gate counts $N_{\mathrm{CNOT}}$ for circuits used to prepare approximate ground states with energies shown in Fig.~\ref{fig:h4plots}(a). For each bond distance, we repeat independent numerical experiments for 10 times. Round points depict experimental runs in which chemical precision was reached, and crosses~-- runs in which it was not reached. We observe that multiobjective search can find compressed circuits with thirty to forty two-qubit gates for the majority of bond distances. Overall we note the gradual increase of required circuit size as $d$ grows. This aligns well with the fact that at large bonds hydrogen chains exhibit strong correlations, as evidenced by large discrepancy between Hartree-Fock energy and FCI energy \cite{Tang2019,QunaSys}. Additionally, we observe a noticeable optimization glitch at around $0.90$~\AA, which hinders efficient parameter research. While the origin of this phenomenon remains unclear, we narrowed down the area and revealed numerically that observed behavior resembles a computational phase transition. 

The progress of variational optimization of circuit topology is demonstrated in Fig.~\ref{fig:h4plots}(c), where Pareto fronts for several generations are shown (we choose $d=1.20$~\AA). While circuits of $N_{\mathrm{CNOT}} = 35$ for reaching chemical accuracy can be prepared already at 19th generation, systematic lowering of two-qubit gate count is difficult, driven by strong interparticle correlations present in the system.

Fig.~\ref{fig:h4plots}(d) shows a comparison of the proposed approach to HEA VQE for the bond distance of 1.20~{\AA}ngstrom, which was run until convergence for increasing number of layers. We observe that the hardware-efficient ansatz requires a minimum of 30 layers (210 CNOTs) to reach chemical precision (and at least 40 layers to reach it with high probability), while the proposed MoG-VQE method produces accurate circuits with as low as 31 CNOTs as seen from ~Fig.~\ref{fig:h4plots}(c).

\textit{Lithium hydride LiH (12 qubits).} Another considered large molecule corresponds to LiH decomposed into qubit form without simplifications \cite{QunaSys} and mapped onto $N = 12$ qubits. The bond distance of $2.00$~{\AA} is considered.
Results of experimental runs are shown in Fig.~\ref{fig:LiH}. In Fig.~\ref{fig:LiH}(a) we present Pareto fronts for several generations produced by MoG-VQE. Starting from compact circuits at early generations, the optimizer can efficiently lower both the energy and the gate count. Overall this corresponds to rather weak correlations for LiH, as was also noted before~\cite{Tang2019}. We compare results to HEA VQE simulation [Fig.~\ref{fig:LiH}(b)] where several independent runs of sep-CMA-ES are performed, yet we do not observe convergence to the chemically precise energy. We observe that at shallow depth circuits suffer from weak expressibility and simply cannot prepare the ground state of the 12-qubit Hamiltonian, while at large circuit depth ($p \geq 25$) training becomes inefficient, leading to energy increase. This is ultimately linked to increased number of optimization parameters and scaling for the CMA-ES optimizer.

As an example of a high-performing circuit for preparing LiH ground state, we provide the discovered circuit scheme with only 12 CNOTs from another experimental run. The circuit is visualized with Qiskit~\cite{Qiskit} and explicitly shown in Supplementary Figure~\ref{fig:LiHcircuit}. We note that this circuit is not optimized, nor it is transpiled to a specific geometry. These steps are however important when considered for real hardware implementation. While two-qubit operations in Fig.~\ref{fig:LiHcircuit} may appear long-range interacting in some geometries, this is highly dependent on actual quantum processor topology. For instance, trapped ion processors and transmon-based processors in the star topology allow implementation with zero overhead~\cite{Song2017}, while modern two-dimensional square grid processors~\cite{Arute2019} can lower it significantly. In particular, when optimization for specific hardware is considered, a separate qubit routing procedure can be implemented via existing compilers~\cite{Cowtan2019} such that the overhead for non-neighbouring two-qubit gates is kept at minimal value.


\section{Discussion}

To highlight the potential yield of the designed multiobjective genetic VQE we can take a simple noise model, where each single- and two-qubit operation is associated to an error (e.g. bit flip), and errors are uncorrelated. Notably, this holds even for large-scale quantum computation, as shown in the quantum supremacy experiment~\cite{Arute2019}. The representative circuit for LiH contains $N_{\mathrm{CNOT}} = 12$ CNOTs and $N_{\mathrm{rot}} = 84$ non-zero single-axis rotations, that can be contracted into $N_{\mathrm{rot}} = 32$ general gates. This is associated to noise levels $\varepsilon_{\mathrm{CNOT}} \sim 10^{-2}$ and $\varepsilon_{\mathrm{rot}} \sim 10^{-3}$, correspondingly \cite{Arute2019,IonQ2019}. Fidelity for the ground state preparation of 12-qubit molecule can be estimated as $F = (1 - \varepsilon_{\mathrm{rot}})^{N_{\mathrm{rot}}} (1 - \varepsilon_{\mathrm{CNOT}})^{N_{\mathrm{CNOT}}}$ and is over $85\%$, making ground state estimation reliable. 

To conclude, we proposed the multiobjective genetic variational quantum eigensolver based on evolutionary algorithms. It utilizes the genetic algorithm NSGA-II to optimize the variational ansatz topology guided by the requirement to minimize the two-qubit gate count. For each scheme we optimize variational angles using CMA-ES evolutionary strategy that proves to work well for black-box functions. The strategy points to the ultimate limit where even usual hardware-efficient layered ansatz is expensive and excessive, as in certain cases only targeted entanglement between qubits might suffice to prepare ground state approximation. The approach is particularly useful for problems where two-qubit operations are noisy and their use needs to be minimized. We tested the approach for different molecular Hamiltonians encoded in 8 and 12 qubits, and found compact schemes which can readily prepare ground state approximations.

Recently, we became aware of two conceptually similar approaches ({\sffamily{}Rotoselect} \cite{rotoselect} and Evolutionary Variational Quantum Eigensolver (EVQE) \cite{EVQE}) having similar workflows with ansatz modification. In {\sffamily{}Rotoselect} the distinct method for choosing generators of rotations was used, thus allowing to reach better energy accuracy with smaller number of layers compared to standard HEA VQE. EVQE uses a different genetic heuristic to modify variational ansatz operating at the level of layers (add/remove), and exploits a single objective function. In the current work we go beyond this approach by using block-structured ansatz and employing multiobjective approach that allows searching for Pareto-optimal solutions.

We shall stress that the proposed algorithm can be further improved. First, the question of optimal gate block choice remains open, and better parametrization can largely yield the performance of the algorithm. The promising approach (yet to be implemented) is building an ansatz from blocks that respect symmetries of molecular systems (for instance, charge conservation)~\cite{Gard2019}. For instance, by choosing a less ``powerful'' block with only single $\hat{Y}$ rotation we can largely facilitate the optimization process, reducing its wall-time, though at the expense of larger circuit depth. Also, recent results suggest that introduced correlation may largely help when searching for optimal VQE parameters~\cite{Volkoff2020}.
Next, different initialization schemes will be considered in the future to further facilitate the convergence. Currently we do not reuse data when optimizing circuits, and continue to optimize all angles in the circuit. One improved strategy may account for quasi-optimal preparatory sequences, and ``freezing'' them to yield further fine-tuned optimization. 
Finally, while multiobjective genetic nature of the algorithm is important for changing ansatz topology, the intermediate variation for optimal angles can be performed by using gradient-based methods with automatic differentiation in the forward mode~\cite{Mitarai2019}, likely improving the convergence.\\

\textit{Acknowledgements.} The authors acknowledge the support from the mega-grant No. 14.Y26.31.0015 of the Ministry of Education and Science of the Russian Federation, the Government of Russian Federation (Grant 08-08), and ITMO Fellowship and Professorship Program.

\clearpage

\setcounter{figure}{0} 
\renewcommand{\thefigure}{S\arabic{figure}}
\renewcommand{\thesection}{S\arabic{section}}
\renewcommand{\theequation}{S\arabic{equation}}

\newpage
\begin{figure*}
\centering
\textbf{Supplementary Circuit: Ground state preparation for LiH}
\includegraphics[width=1.\linewidth]{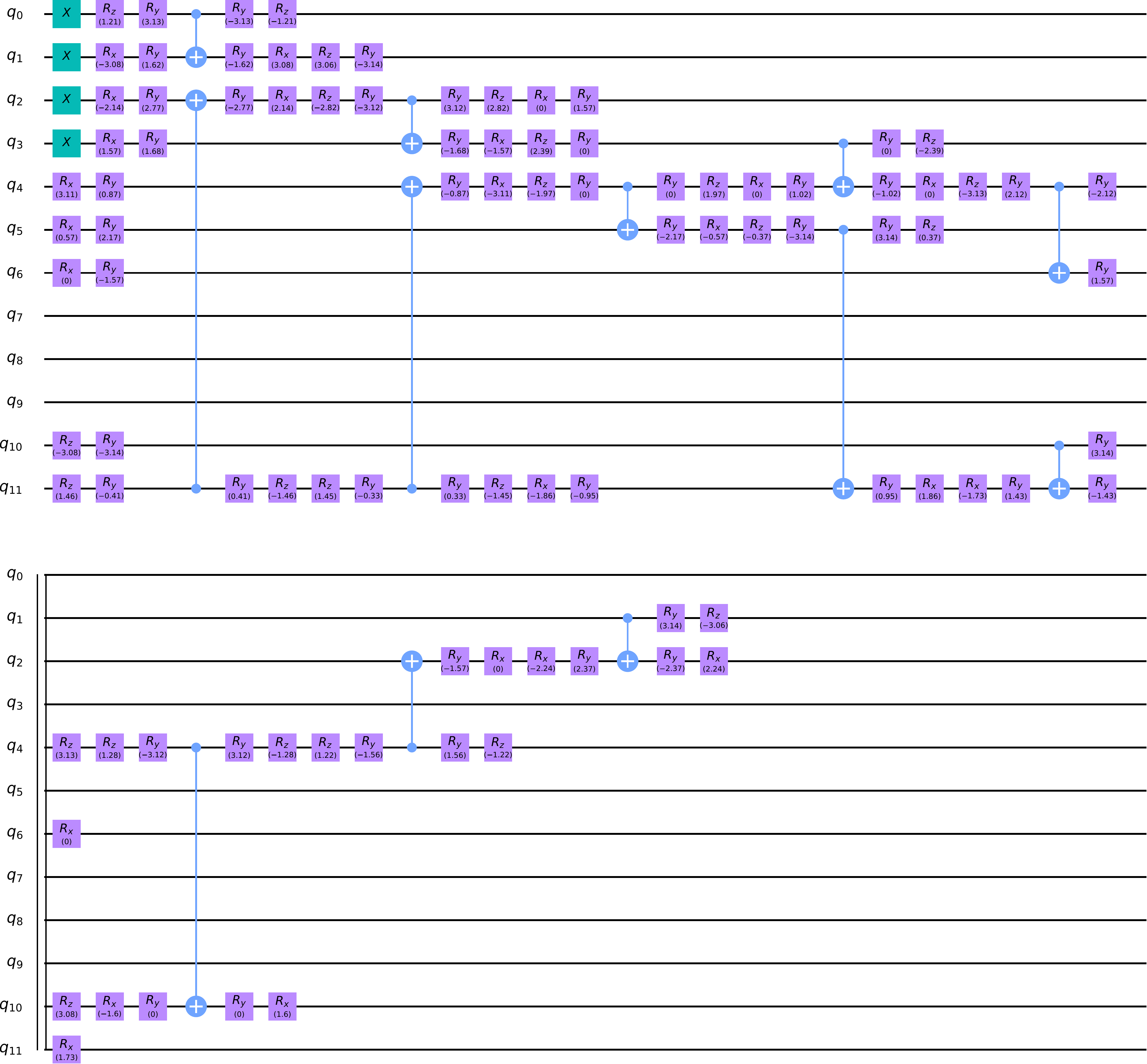}
\caption{Ground state preparation circuit for LiH molecule described by $N=12$ qubits, found by the developed MoG-VQE algorithm. Here an unoptimized platform-agnostic version is shown, which can be further compressed to reduce the single-qubit gate count when more rotations are considered. Blue gates are CNOTs, and purple blocks are rotations generated by $\hat{X}$, $\hat{Y}$, $\hat{Z}$.}
\label{fig:LiHcircuit}
\end{figure*}


\begin{thebibliography}{99}

\bibitem{OxfordRev} S. McArdle, S. Endo, A. Aspuru-Guzik, S. Benjamin, and Xiao Yuan, \textit{Quantum computational chemistry}, Rev. Mod. Phys. \textbf{92}, 15003 (2020).

\bibitem{ZapataRev} Y. Cao, J. Romero, J. P. Olson, M. Degroote, P. D. Johnson, M. Kieferov\'{a}, I. D. Kivlichan, T. Menke, B. Peropadre, N. P. D. Sawaya, Sukin Sim, L. Veis, and A. Aspuru-Guzik, \textit{Quantum Chemistry in the Age of Quantum Computing}, Chem. Rev. \textbf{119}, 10856-10915 (2019); arXiv:1812.09976.

\bibitem{Reiher2016} M. Reiher, N. Wiebe, K. M Svore, D. Wecker, and M. Troyer, \textit{Elucidating Reaction Mechanisms on Quantum Computers}, PNAS \textbf{114}, 7555 (2017).

\bibitem{Babbush2019} R. Babbush, D. W. Berry, J. R. McClean, and H. Neven, \textit{Quantum simulation of chemistry with sublinear scaling in basis size}, npj Quantum Information \textbf{5}, 92 (2019).


\bibitem{Lloyd1996} S. Lloyd, \textit{Universal Quantum Simulators}, Science
\textbf{273}, 1073-1078 (1996).

\bibitem{Aspuru-Guzik2005} A. Aspuru-Guzik, A. D. Dutoi, P. J. Love, M. Head-Gordon, \textit{Simulated Quantum Computation of Molecular Energies}, Science \textbf{309}, 1704 (2005).

\bibitem{Preskill2018} J. Preskill, \textit{Quantum Computing in the NISQ era and beyond}, Quantum \textbf{2}, 79 (2018); arXiv:1801.00862.

\bibitem{Peruzzo2014} A. Peruzzo, J. McClean, P. Shadbolt, M.-H. Yung, X.-Q. Zhou, P. J. Love, A. Aspuru-Guzik, and J. L. O'Brien, \textit{A variational eigenvalue solver on a photonic quantum processor}, Nature Comm. \textbf{5}, 4213 (2014).

\bibitem{McClean2016} J. McClean, J. Romero, R. Babbush, and A. Aspuru-Guzik, \textit{The theory of variational hybrid quantum-classical algorithms}, New J. Phys. \textbf{18}, 023023 (2016).

\bibitem{Romero2019} J. Romero, R. Babbush, J. R. McClean, C. Hempel, P. Love, and A. Aspuru-Guzik, \textit{Strategies for quantum computing molecular energies using the unitary coupled cluster ansatz}, Quantum Sci. Technol. \textbf{4}, 014008 (2019).


\bibitem{OMalley2016} P. J. J. O'Malley \textit{et al.}, \textit{Scalable Quantum Simulation of Molecular Energies}, Phys. Rev. X \textbf{6}, 031007 (2016).

\bibitem{IonQ2019} Y. Nam, J.-S. Chen, N. C. Pisenti, K. Wright, C. Delaney, D. Maslov, K. R. Brown, S. Allen, J. M. Amini, J. Apisdorf, K. M. Beck, A. Blinov, V. Chaplin, Mika Chmielewski, C.Collins, S. Debnath, A. M. Ducore, K. M. Hudek, M. Keesan, S. M. Kreikemeier, J. Mizrahi, P. Solomon, M. Williams, J. D. Wong-Campos, C. Monroe, J. Kim, \textit{Ground-state energy estimation of the water molecule on a trapped ion quantum computer}, npj Quantum Information \textbf{6}, 33 (2020).

\bibitem{Kandala2017} A. Kandala, A. Mezzacapo, K. Temme, M. Takita, M. Brink, J. M. Chow, and J. M. Gambetta, \textit{Hardware-efficient variational quantum eigensolver for small molecules and quantum magnets}, Nature (London) \textbf{549}, 242 (2017).

\bibitem{Ganzhorn2018} M. Ganzhorn, D. J. Egger, P. Kl. Barkoutsos, P. Ollitrault, G. Salis, N. Moll, A. Fuhrer, P. Mueller, S. Woerner, I. Tavernelli, and S. Filipp, \textit{Gate-efficient simulation of molecular eigenstates on a quantum computer}, Phys. Rev. Applied \textbf{11}, 044092 (2019).

\bibitem{Hempel2018} C. Hempel, C. Maier, J. Romero, J. McClean, T. Monz, H. Shen, P. Jurcevic, B. P. Lanyon, P. Love, R. Babbush, A. Aspuru-Guzik, R. Blatt, and C. F. Roos, \textit{Quantum Chemistry Calculations on a Trapped-Ion Quantum Simulator}, Phys. Rev. X \textbf{8}, 031022 (2018).

\bibitem{Arute2020} F. Arute et al., \textit{Hartree-Fock on a superconducting qubit quantum computer}, arXiv:2004.04174 (2020).

\bibitem{McArdle2018} S. McArdle, T. Jones, S. Endo, Ying Li, S. Benjamin, and Xiao Yuan, \textit{Variational ansatz-based quantum simulation of imaginary time evolution}, npj Quantum Information \textbf{5}, 75 (2019).

\bibitem{Endo2018} S. Endo, T. Jones, S. McArdle, Xiao Yuan, and S. Benjamin, \textit{Variational quantum algorithms for discovering Hamiltonian spectra}, arXiv:1806.05707 (2018).

\bibitem{Stokes2020} J. Stokes, J. Izaac, N. Killoran, and G. Carleo, \textit{Quantum Natural Gradient}, Quantum \textbf{4}, 269 (2020).

\bibitem{Yamamoto2019} N. Yamamoto, \textit{On the natural gradient for variational quantum eigensolver}, arXiv:1909.05074 (2019).

\bibitem{Balint2019} B. Koczor and S. C. Benjamin, \textit{Quantum natural gradient generalised to non-unitary circuits}, arXiv:1912.08660 (2019).


\bibitem{Ryabinkin2019} I. G. Ryabinkin, S. N. Genin, and A. F. Izmaylov, \textit{Constrained Variational Quantum Eigensolver: Quantum Computer Search Engine in the Fock Space}, J. Chem. Theory. Comput. \textbf{15}, 249 (2019).

\bibitem{Gard2019} B. T. Gard, L. Zhu, G. S. Barron, N. J. Mayhall, S. E. Economou, E. Barnes, \textit{Efficient symmetry-preserving state preparation circuits for the variational quantum eigensolver algorithm}, npj Quantum Information \textbf{6}, 10 (2020).

\bibitem{Riverlane2019} D. Wang, O. Higgott, and S. Brierley, \textit{Accelerated Variational Quantum Eigensolver}, Phys. Rev. Lett. \textbf{122}, 140504 (2019).


\bibitem{Mitarai2019} K. Mitarai, Y. O. Nakagawa, W. Mizukami, \textit{Theory of analytical energy derivatives for the variational quantum eigensolver}, Phys. Rev. Research 2, 013129 (2020).

\bibitem{OBrien2019} T. E. O’Brien, B. Senjean, R. Sagastizabal, X. Bonet-Monroig, A. Dutkiewicz, F. Buda, L. DiCarlo, and L. Visscher, \textit{Calculating energy derivatives for quantum chemistry on a quantum computer}, npj Quantum Information \textbf{5}, 113 (2019).


\bibitem{Cerezo2020} M. Cerezo, A. Sone, T. Volkoff, L. Cincio, and P. J. Coles, \textit{Cost-Function-Dependent Barren Plateaus in Shallow Quantum Neural Networks}, arXiv:2001.00550 (2020).

\bibitem{Kyriienko2019} O. Kyriienko, \textit{Quantum inverse iteration algorithm for programmable quantum simulators}, npj Quantum Information 6, 7 (2020).

\bibitem{Stair2020} N. H. Stair, R. Huang, and F. A. Evangelista, \textit{A Multireference Quantum Krylov Algorithm for Strongly Correlated Electrons}, J. Chem. Theory Comput. 16, 2236 (2020).

\bibitem{Parrish2019} R. M. Parrish and P. L. McMahon, \textit{Quantum Filter Diagonalization: Quantum Eigendecomposition without Full Quantum Phase Estimation}, arXiv:1909.08925 (2019).

\bibitem{Wei2020} S. Wei, H. Li, and G. Long, \textit{A Full Quantum Eigensolver for Quantum Chemistry Simulations}, Research 2020, 1486935 (2020).

\bibitem{Huggins2019} W. J. Huggins, J. Lee, U. Baek, B. O'Gorman, and K. B. Whaley, \textit{A Non-Orthogonal Variational Quantum Eigensolver}, New J. Phys. in press (2020); https://doi.org/10.1088/1367-2630/ab867b.

\bibitem{Grimsley2019} H. R. Grimsley, S. E. Economou, E. Barnes, N. J. Mayhall, \textit{An adaptive variational algorithm for exact molecular simulations on a quantum computer}, Nature Commun. \textbf{10}, 3007 (2019).

\bibitem{Tang2019} H. L. Tang, V. O. Shkolnikov, G. S. Barron, H. R. Grimsley, N. J. Mayhall, E. Barnes, S. E. Economou, \textit{qubit-ADAPT-VQE: An adaptive algorithm for constructing hardware-efficient ansatze on a quantum processor}, arXiv:1911.10205 (2019).



\bibitem{Ryabinkin2019b} I. G. Ryabinkin, T.-C. Yen, S. N. Genin, and A. F. Izmaylov, \textit{Qubit Coupled Cluster Method: A Systematic Approach to Quantum Chemistry on a Quantum Computer}, J. Chem. Theory Comput. \textbf{14}, 6317 (2019).

\bibitem{Herasymenko2019} Y. Herasymenko and T. E. O'Brien, \textit{A diagrammatic approach to variational quantum ansatz construction}, arXiv:1907.08157 (2019).

\bibitem{Lamata2020} F. Albarr\'{a}n-Arriagada, J. C. Retamal, E. Solano, and L. Lamata, \textit{Reinforcement learning for semi-autonomous approximate quantum eigensolver}, Mach. Learn.: Sci. Technol. \textbf{1}, 015002 (2020).

\bibitem{Woitzik2020} A. J. C. Woitzik, P. Kl. Barkoutsos, F. Wudarski, A.Buchleitner, and I. Tavernelli, \textit{Entanglement Production and Convergence Properties of the Variational Quantum Eigensolver}, arXiv:2003.12490 (2020).


\bibitem{mitchell} M. Mitchell, \textit{An Introduction to Genetic Algorithms} (MIT Press, 1996).

\bibitem{nsga} K. Deb, A. Pratap, S. Agarwal and T. Meyarivan, \textit{A fast and elitist multiobjective genetic algorithm: NSGA-II}, IEEE Transactions on Evolutionary Computation \textbf{6(2)}, 182-197, (2002).

\bibitem{Iten2016} R. Iten, R. Colbeck, I. Kukuljan, J. Home, and M. Christandl, \textit{Quantum circuits for isometries}, 
Phys. Rev. A \textbf{93}, 032318 (2016).

\bibitem{NielsenChuang} M. A. Nielsen and I. L. Chuang, \textit{Quantum Computation and Quantum Information}, (Cambridge Univ. Press, 2010).

\bibitem{Grant2019} E. Grant, L. Wossnig, M. Ostaszewski, and M. Benedetti, \textit{An initialization strategy for addressing barren plateaus in parametrized quantum circuits}, Quantum 3, 214 (2019).


\bibitem{bfgs} R. Fletcher, \textit{Practical methods of optimization} (2nd ed.), New York: John Wiley \& Sons (1987).

\bibitem{nelder-mead}  J. A. Nelder and R. Mead, \textit{A Simplex Method for Function Minimization}, The Computer Journal, \textbf{7}, 308 (1965).

\bibitem{adam} D. P. Kingma and J. Ba, \textit{Adam: A method for stochastic optimization}, in Proceedings of the 3rd International Conference on Learning Representations (ICLR) (2015); arXiv:1412.6980.

\bibitem{Kubler2020} J. M. K\"{u}bler, A. Arrasmith, L. Cincio, P. J. Coles, \textit{An Adaptive Optimizer for Measurement-Frugal Variational Algorithms}, Quantum \textbf{4}, 263 (2020).

\bibitem{cmaes} N. Hansen, S. D. M\"{u}ller, P. Koumoutsakos, Evolutionary computation \textbf{11}, 1-18, (2003).

\bibitem{deap} F.-A. Fortin, F.-M. De Rainville, M.-A. Gardner, M. Parizeau and C. Gagn\'{e}, \textit{DEAP: Evolutionary Algorithms Made Easy}, Journal of Machine Learning Research \textbf{13},  2171-2175, (2012).

\bibitem{libcmaes} See online at \url{https://github.com/beniz/libcmaes}

\bibitem{Qutip} J. R. Johansson, P. D. Nation, and F. Nori, \textit{QuTiP 2: A Python framework for the dynamics of open quantum systems}, Comp. Phys. Comm. \textbf{184}, 1234 (2013) [DOI: 10.1016/j.cpc.2012.11.019].

\bibitem{Qiskit} H. Abraham et al., \textit{Qiskit: An Open-source Framework for Quantum Computing}, (2019); \url{https://github.com/Qiskit/qiskit}; doi:10.5281/zenodo.2562110.

\bibitem{QuEST} T. Jones, A. Brown, I. Bush, and S. C. Benjamin, \textit{QuEST and High Performance Simulation of Quantum Computers}, Sci. Rep. \textbf{9}, 10736 (2019); \url{https://quest.qtechtheory.org/}.

\bibitem{psi4} R. M. Parrish \textit{et al.}, \textit{Psi4 1.1: An Open-Source Electronic Structure Program Emphasizing Automation, Advanced Libraries, and Interoperability}, J. Chem. Theory Comput. \textbf{13}, 3185 (2017).

\bibitem{OpenFermion} J. R. McClean \textit{et al.}, \textit{OpenFermion: the electronic structure package for quantum computers}, Quantum Sci. Technol. \textbf{5}, 034014 (2020).



\bibitem{QunaSys} See online at \url{https://quantaggle.com/competitions/gs-pes/} and  \url{https://github.com/qulacs/Quantaggle_dataset/tree/master/datasets/Small_Molecules_1}.

\bibitem{PySCF} Q. Sun, T. C. Berkelbach, N. S. Blunt, G. H. Booth, S. Guo, Z. Li, J. Liu, J. McClain, E. R. Sayfutyarova, S. Sharma, S. Wouters, and G. K.-L. Chan, \textit{PySCF: the Python‐based simulations of chemistry framework}, WIREs Comput. Mol. Sci. \textbf{8}, e1340 (2018); doi:10.1002/wcms.1340


\bibitem{sepCMAES} R. Ros and N. Hansen, \textit{A Simple Modification in CMA-ES Achieving Linear Time and Space Complexity}, Parallel Problem Solving from Nature, Lecture Notes in Computer Science \textbf{5199}, 296-305 (2008). 

\bibitem{Song2017} C. Song, K. Xu, W. Liu, C. Yang, S.-B. Zheng, H. Deng, Q. Xie, K. Huang, Q. Guo, L. Zhang, P. Zhang, D. Xu, D. Zheng, X. Zhu, H. Wang, Y.-A. Chen, C.-Y. Lu, S. Han, and J.-W. Pan, \textit{10-Qubit Entanglement and Parallel Logic Operations with a Superconducting Circuit}, Phys. Rev. Lett. \textbf{119}, 180511 (2017).

\bibitem{Arute2019} F. Arute et al., \textit{Quantum supremacy using a programmable superconducting processor}, Nature \textbf{574}, 505-510 (2019).


\bibitem{rotoselect} M. Ostaszewski, E. Grant, and M. Benedetti, \textit{Quantum circuit structure learning}, arXiv:1905.09692 (2019).

\bibitem{EVQE} A. G. Rattew, S. Hu, M. Pistoia, R. Chen, and S. Wood, A \textit{Domain-agnostic, Noise-resistant, Hardware-efficient Evolutionary Variational Quantum Eigensolver}, arXiv:1910.09694 (2019).


\bibitem{Cowtan2019} A. Cowtan, S. Dilkes, R. Duncan, A. Krajenbrink, W. Simmons, S. Sivarajah, \textit{On the qubit routing problem}, Leibniz International Proceedings in Informatics (LIPIcs), doi:10.4230/LIPIcs.TQC.2019.5 (2019); arXiv:1902.08091.

\bibitem{Volkoff2020} T. Volkoff and P. J. Coles, \textit{Large gradients via correlation in random parameterized quantum circuits}, arXiv:2005.12200 (2020).


\end{thebibliography}
\end{document}